\documentclass[aps,prb,twocolumn]{revtex4}
\usepackage{graphicx}
\usepackage{amssymb}%
\usepackage{amsmath}
\usepackage{wasysym}
\usepackage{ulem}
\usepackage{color}

\begin{document}
\title{Resonating color state and emergent chromodynamics in the kagome antiferromagnet}
\author{O.~C\'epas and A.~Ralko}
\affiliation{Institut N\'eel, CNRS et Universit\'e Joseph Fourier, BP 166, F-38042 Grenoble cedex 9, France }

\begin{abstract}
We argue that the spin-wave breakdown in the Heisenberg kagome
antiferromagnet signals an instability of the ground state and leads, through an emergent local constraint, to a quantum dynamics described by a
gauge theory similar to that of chromodynamics. For integer spins, we
show that the quantum fluctuations of the gauge modes select the
$\sqrt{3} \times \sqrt{3}$ N\'eel state with an on-site moment
renormalized by color resonances.  We find non-magnetic low-energy
excitations that may be responsible for a deconfinement ``transition''
at experimentally accessible temperatures which we estimate.
\end{abstract}

\maketitle

Spin liquids are intriguing states of matter that do not  break any symmetry at zero temperature, and were conjectured by Anderson in the form of a 
 resonating valence bond (RVB) liquid.\cite{Anderson} Contrary to conventional magnets which have long-range order and spin wave excitations described by an effective non-linear sigma model,\cite{Chakra} RVB states have emergent gauge excitations.~\cite{Baskaran} Gauge degrees of freedom appear because of local constraints within a low-energy manifold, such as the hard-core constraint for singlets\cite{Baskaran} or, more generally, some ``ice-rules''.\cite{Gingras} As a result, the magnetism can be described in terms of ``electrodynamics'' and Coulomb phases which find striking experimental realizations in spin-ice systems at finite temperatures.\cite{Gingras} Recently, experiments on the two-dimensional spin-1/2 Heisenberg
kagome compound ZnCu$_3$(OH)$_6$Cl$_3$ suggest a possible quantum spin-liquid
state at zero temperature.\cite{Mendels} Theoretically for $S=1/2$ systems, there is indeed evidence for a spin-liquid state,\cite{Yan} not inconsistent\cite{Lhuillier,Sakai} with a gapless RVB state.\cite{Ran}
From a different route, the large-S approach provides important insights, such as integer \textit{vs}.  half-integer spin
effects.\cite{Haldane} For the kagome system, the large-$S$ spin-wave theory
is known to break down because of  \textit{local}
(weathervane) modes with diverging fluctuations.\cite{Harris,Chalker,Ritchey} This is the consequence of a macroscopic number of classically degenerate states that can be seen as the 3-colorings of the kagome lattice with spins pointing at 120$^o$ apart (Fig. 1).
Contrary to spin-ice models where the constraint is enforced by strong local anisotropies,\cite{Gingras} here the local (color) constraint arises dynamically because of an order-by-disorder mechanism.\cite{Harris,Chalker,Ritchey} In this paper, we show that the spin-wave breakdown is the signature of emergent gauge degrees of freedom that are governed by a Hamiltonian similar to that of quantum chromodynamics. The issue is whether the present theory stabilizes a resonating color ``crystal'' (with a finite N\'eel order-parameter and color resonances) or leads to a delocalization in the low-energy manifold, thus realizing a resonating color ``spin-liquid''  (instead of RVB).

\begin{figure}[h]
\includegraphics[width=0.45\textwidth,clip]{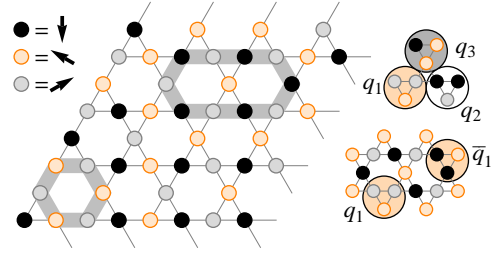}
\caption{(color online.) Loop-induced low-energy dynamics within the 3-coloring manifold of the kagome lattice (spin/color equivalence is shown), described as ``gluon'' dynamics. Right: the $q_1\bar{q}_1$ (``meson'') and $q_1q_2q_3$ (``baryon'') are high-energy ``matter'' excitations violating the local constraint.}
\label{fig00}
\end{figure}

First, we derive the effective lattice gauge theory in the low-energy
manifold of 3-colorings starting from large-$S$. It is known both from
classical Monte Carlo simulations ($S=\infty$)\cite{Chalker,Reimers}
and from exact spectra at finite $S$\cite{Rousochatzakis} that the
3-coloring states (spin-ice subspace) are relevant at low energy. To remain in the low-energy manifold, single spin-flips
are not allowed and the spin motion consists of the tunneling of a
collective loop of spins across an energy barrier separating
degenerate states.\cite{Delft} The loops are closed loops of $L$
spins of two colors (Fig. 1).  Classically, they can be rotated freely
about the effective field of the third-color spins, for any arbitrary
collective angle $\phi$.  However, this local rotation can be viewed
as a defect for the spin-waves and, therefore, increases the
zero-point energy. In the Born-Oppenheimer approximation, this
generates an effective energy barrier for the slow loop
motion.\cite{Ritchey,Doucot}  We now calculate the energy barrier
starting with the quantum Heisenberg Hamiltonian
\begin{equation}
H= J \sum_{<i,j>} S_i\cdot S_j,
\label{H}
\end{equation}
with $J$ the coupling between nearest-neighbor spins $S$. 
To do so, we start with a classical ground state and consider a given loop (see Fig. 1 for example) rotated by $\phi$. The Holstein-Primakov spin-wave equations of motion are set up in real space (given that translation invariance is broken) and diagonalized for different $\phi$ (up to $N=972$ sites). We compute the zero-point energy
\begin{figure}
\includegraphics[width=0.45\textwidth,clip]{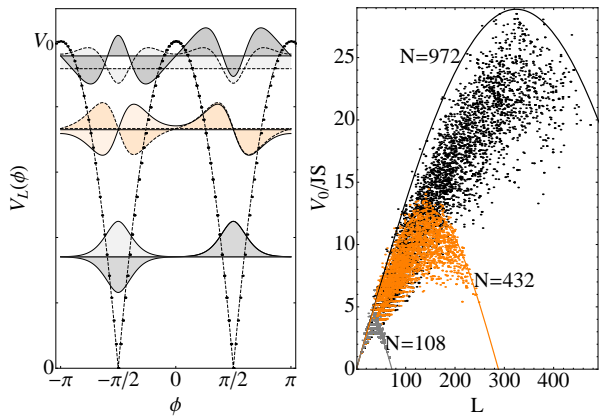}
\caption{(color online.) Left: Double-well potential for the loop motion and quantized levels with even (solid line) and odd (dashed line) wavefunctions. Right: Barrier height distribution as a function of loop length in the 3-coloring manifold. The curves shown are simple cosine functions.}\label{fig0}
\end{figure}
\begin{equation}
V_L(\phi) =  \sum_{l=1}^N \frac{1}{2} \omega_l(\phi),
\label{potential}
\end{equation}
where $l$ runs over all spin-wave energies, $\omega_l$. It is a double-well
potential with two minima located at $\pm \pi/2$ (Fig.~\ref{fig0}) corresponding indeed to the selection of 3-colorings. The potential is well fitted by an inverted parabolic function, which we will use in the following.
Repeating the calculation for loops of arbitrary lengths in a Monte Carlo
sampling of the manifold, we have systematically computed the barrier heights $V_0$  (Fig.~\ref{fig0}). For small
loops, we find $V_0=\eta JSL$ with $\eta=0.14$, in agreement with
Ref.~\onlinecite{Ritchey}; the barriers are rather small. For larger loops, a dispersion about the mean value is
observed, which tells us that the barrier heights depend on the configuration. Note that the symmetry about $L=N/3$ occurs because rotating a loop of $L=2N/3$ results in choosing another global plane at no energy cost.   We also point out that anharmonic corrections will give additional
contributions $\sim JS^{2/3}$.\cite{Chubukov,Henley2,Delft}  The kinetic energy of a loop is obtained by
considering the spins of the loop in the effective field of the fixed third color spins (say along $-z$). The Goldstone mode corresponds to the collective rotation about $z$ and is taken separatly.\cite{Trumper} Together with $V_L$ we
obtain\cite{CR} the hamiltonian for a single loop
\begin{equation}
H_L=\frac{1}{2\chi}\left(\hat{S}^z-A\right)^2 + V_L(\phi)
\label{HL0}
\end{equation}
where $\hat{S}^z=\sum_{i=1}^L S_i^z=-i \partial
/\partial \phi$ is the total spin of the loop, conjugate to $\phi$,
and $\chi =L/(4J)$ the moment of inertia.  $H_L$ is similar to an
Aharonov-Bohm ring in a double-well potential. The flux $A=LS/2$ is
the classical magnetization of the loop. For the kagome lattice, the
only possible loops have $L=4n+2$ ($n$ an integer),\cite{Delft} so
that $A=(2n+1)S$.  By considering the Berry phase $2\pi A$, von Delft
and Henley predicted a destructive interference for half-integer
$S$.\cite{Delft} Indeed, the first term of (\ref{HL0}) is minimized for 
$S^z=A$ if $A$ is an integer or $S^z=A \pm 1/2$ if $A$ is half-integer (``Kramers'' doublet).  In the presence of the external potential,
this remains true.  By using a simple gauge transformation $\Psi(\phi)
= e^{-i A \phi} u(\phi)$, $u(\phi)$ satisfies the Schr\"odinger
equation with either periodic (integer $S$) or antiperiodic
(half-integer $S$) boundary conditions. $V_L(\phi)$ has a periodicity
of $\pi$ so we expect Bloch states with energies $E_k$. Periodic
(resp. anti-periodic) boundary conditions selects the bonding $k=0$
and anti-bonding $k=1$ states (resp.~$k=\pm 1/2$), which have energy
difference $t_L$. Given that
$V_L(\phi)=V_L(-\phi)$, we have $E_{-k}=E_k$ and the two states $k=\pm
1/2$ remains degenerate ($t_L=0$).

To extract $t_L$, we 
  solve  exactly the Schr\"odinger equation in the double-well
  parabolic potential with parabolic cylinder
  wavefunctions (Fig.~\ref{fig0}).  In the
  semi-classical limit (large $S$ or large $L$), the solution
  coincides with the Wentzel-Kramers-Brillouin approximation and we obtain
\begin{eqnarray}
\begin{cases}
 t_L = a J \exp \left( -\frac{\pi^2}{4} \sqrt{\eta S/2} L  \right) ~~~~\textrm{(integer S)}  \\
t_L = 0 ~~~~~~~~~~~~~~~~~~~~~~~~~~ \textrm{(half-integer S)}  \label{t}
\end{cases} 
\end{eqnarray}
where $a=0.261(\eta S)^{11/12}L^{5/6}$ and taking $\eta=V_0/JSL$, i.e. neglecting the dispersion of the barrier heigths. Quantum tunneling of loops of size $L>L_c= 4/\pi^2\sqrt{\eta S/2}$ is therefore supressed. 

We now consider the resulting low-energy dynamics on the lattice. In this derivation, the lowest energy states are the 3-coloring quantum coherent states 
$|C\rangle$.\cite{ortho} The number of
states scales like $W \sim 1.13^N$ in the thermodynamic
limit\cite{Baxter,Jesper} and is much smaller than that of the
original spin problem (\ref{H}), $(2S+1)^N$.  The dynamics of the loops is described by the quantum Hamiltonian
\begin{equation}
H_e=-\sum_{C,C'} t_{C,C'} |C\rangle \langle C'| + \sum_C  E_{C} |C\rangle \langle C|
\label{hamm00}
\end{equation}
where $t_{C,C'}=t_L$ if the two states $|C \rangle$ and $|C' \rangle$
are connected by the tunneling of a loop of size $L$ and $t_{C,C'}=0$
otherwise. The model depends only on $S$ and the largest energy scale
is $t_6 \equiv t$.  Anharmonicity~\cite{Chubukov,Henley2} and further neighbor
couplings generate a finite $E_C$ and phase competitions that are beyond the scope of the paper\cite{CR} (here $E_C=0$).
To rewrite $H_e$ exactly in terms of exchange operators of three colors, we have to use the $3\times 3$ Gell-Mann matrices  $\lambda_{c=1,2,3}^{\pm}$ of the SU(3) algebra. We find
\begin{equation}
H_e = -t_6  \sum_{\hexagon} \sum_{c=1}^3  \lambda_{c,1}^+ \lambda_{c,2}^-\lambda_{c,3}^+ \lambda_{c,4}^-\lambda_{c,5}^+ \lambda_{c,6}^-  + h.c. +\cdots 
\label{hamm}
\end{equation}
where the dots stand for longer loops. The Hamiltonian (\ref{hamm}) is
a pure lattice gauge model describing the loop dynamics, and is
similar to that of quantum chromodynamics with the local SU(3)
symmetry broken to U(1)$\times$U(1).\cite{gauge} The weathervane modes
can therefore be seen as ``gluons''. Note that the use of the eight
operators of SU(3) is necessary because here, contrary to
Ref.~\onlinecite{Xu}, the Hamiltonian respects the global Z$_3$
symmetry and flips the three types of loops. If we pursue the analogy,
the finite-energy excitations are defects in the ice rule that consist
of (i) flipping two colors along a finite string,\cite{Kondev} it can
be seen as a ``quark-anti-quark'' $q_i\bar{q}_i$ (meson)
(Fig~\ref{fig00}). It terms of spin, it is a non-magnetic pair of
domain-walls with irrational magnetization
$\sqrt{3}(+\hat{e}_i,-\hat{e}_i)$ ($\hat{e}_i$ are three unit-vectors
with $\hat{e}_1+\hat{e}_2+\hat{e}_3=0$).  (ii) permuting a triangle,
this generates three quarks in a color singlet $q_1q_2q_3$ or
$\bar{q}_1 \bar{q}_2 \bar{q}_3$ (baryon), or, in the spin language, a
non-magnetic excitation
$\sqrt{3}(\hat{e}_1,\hat{e}_2,\hat{e}_3)$. These are exactly the first
excitations of the Potts model, whereas there are
other excitations for the Heisenberg model.  We will address below the
issue of confinement of these excitations. Finally, we point out that
$H_e$ has a complex topological structure separated in
Kempe~\cite{Mohar} and topological sectors.\cite{Castelnovo} While
topological sectors can be connected only by flipping \textit{winding}
loops, Kempe sectors are ensembles of topological sectors that cannot
be connected by such global moves.\cite{Mohar,Huse}
\begin{figure}[h]
\includegraphics[width=0.45\textwidth,clip]{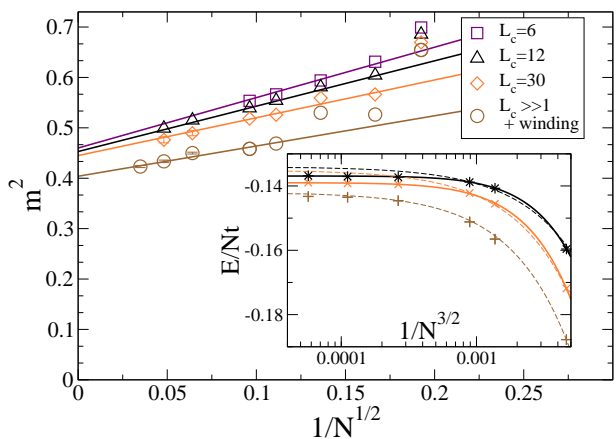}
\caption{(color online.) Finite-size scaling of the order-parameter of the $\sqrt{3} \times \sqrt{3}$ N\'eel state, extrapolating to a finite value (exact diagonalization and quantum Monte Carlo). Inset: total energy and fits assuming gapped (solid lines) or gapless (dashed lines) excitations.}
\label{opfss}
\end{figure}

\begin{figure}[h]
\includegraphics[width=0.45\textwidth,clip]{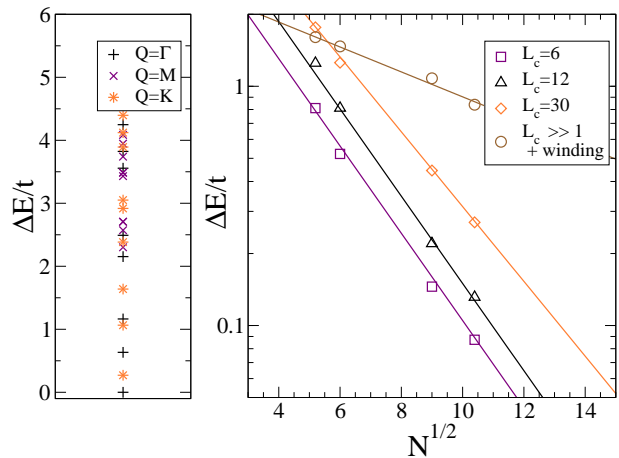}
\caption{(color online.) Left: energy spectrum ($N=108$) showing no clear separation of scale. Right: the lowest \textit{finite-size} gap follows $\Delta E/t \sim \exp(-\alpha N^{1/2})$ confirming the broken symmetry of the $\sqrt{3} \times \sqrt{3}$ N\'eel order.}
\label{gap}
\end{figure}

The ground state is in general a resonating color state, $\Psi =
\sum_C A_C |C \rangle$ in a given sector and the issue is whether it breaks the Z$_3$ symmetry or not. There is an
extension of the model for which $\Psi$ is known exactly.  By adding
$E_C=U_L n_L(C)$ which penalizes states with a large number $n_L(C)$
of loops of size $L$, one obtains a Rokhsar-Kivelson (RK) model for
color dimers.\cite{Castelnovo} In this case, the ground state is the resonating color ``spin-liquid'', $\Psi=\sum_C |C\rangle$ in each
topological sector at the RK point, $U_L=t_{L}$.\cite{Castelnovo}
We study the model (\ref{hamm}) with $E_C=0$ by Lanczos exact
diagonalization of clusters of size
$N=27,36,54,81,108$ and $T=0$ quantum Monte Carlo
(up to $N=675$), since $H_e$ is free of the sign problem.  We find that
(i) $\Psi$ belongs to the largest topological and Kempe sector (which
is non degenerate and has all topological numbers equal to $l/3$ where
$l$ is the linear dimension).  (ii) $\Psi$ has larger amplitudes onto
the six $\sqrt{3} \times \sqrt{3}$ states, and
a finite associated N\'eel order-parameter, $m^2$ which extrapolates to a finite value for infinite size, $m_{\infty} \sim 0.63-0.68$, depending weakly on $L_c$ (Fig.~\ref{opfss}).
 To confirm the N\'eel order, we also calculate the low-energy spectrum by exact diagonalization.  For such a discrete broken symmetry, we expect six quasi-degenerate states separated from higher energy states. While there is no clear separation of scale, the gap does decrease as $\exp(-\alpha N^{1/2})$ which reflects the development of a macroscopic energy barrier (Fig.~\ref{gap}). The low-energy spectrum is therefore perfectly compatible with a N\'eel order with the same pattern as predicted earlier~\cite{Chubukov,Henley2,Sachdev} and on-site magnetization renormalized by color fluctuations. However, there are unconventional non-magnetic excitations at low-energy of order $t$ (gluons) which bear some similarity with the low-lying states of exact spectra.\cite{Lecheminant,Rousochatzakis} 

To discuss the consequences of these unconventional excitations and analyze these results more quantitatively, we consider the effective gauge theory in the continuum limit. The ice-rule provides an exact mapping of the discrete states onto two-component height vectors $\mathbf{h}(r)$ living on the dual lattice.\cite{Huse,Kondev,Korshunov} For instance the six $\sqrt{3} \times \sqrt{3}$ states map onto flat interfaces. 
At $T=0$, the simplest action compatible with the symmetries is a sine-Gordon model,
\begin{equation}
S = \frac{K}{2c^2}\int d^2 \mathbf{x}  d\tau \left[  (\partial_{\tau} \mathbf{h})^2+ c^2 (\nabla \mathbf{h})^2 + \frac{\Delta^2}{2\pi^2} \sum_{\alpha=1}^3 \cos(\mathbf{Q}_{\alpha}.\mathbf{h} )  \right]
\label{action}
\end{equation}
where $K$ is the stiffness, $\Delta^2$ the amplitude of  the locking
potential, and $|\mathbf{Q}_{\alpha}|=4\pi/\sqrt{3}$.\cite{Huse,Kondev,Korshunov} Classical minima of (\ref{action}) correspond to the ``flat'' $\sqrt{3} \times \sqrt{3}$ N\'eel states, with zero-point oscillations described by the two ``gluon'' modes $\omega_{\mathbf{q}}=(c^2\mathbf{q}^2+\Delta^2)^{1/2}$.
 At $T=0$ and in (2+1) dimensions, no
quantum phase transition is expected.\cite{Kosterlitz} The gluons mediate an interaction between the defects (quarks) that is confining if $\Delta$ is finite as $V(r)\sim \Delta r$.\cite{Polyakov} To obtain these parameters, we have used the finite-size
correction to the total energy,  $ \frac{E}{N}=e_0
-0.6586 \frac{2c}{N^{3/2}}$ for a hexagonal lattice of $N$ sites, without a gap (dashed line in Fig.~\ref{gap}). We clearly see that the fit is improved at large distance (for a finite $L_c$) by using $\frac{1}{N^{3/2}} \exp(-l/\xi)$ where $\xi=c/\Delta$ gives the confinement scale (solid line), and $c=13.4t$, $\xi=15.5$,
$e_0=-0.1430t$.  To estimate the stiffness we calculate the
order-parameter, $
m_{\infty}= \langle e^{i\mathbf{Q}.\mathbf{h}} \rangle$ with $|\mathbf{Q}|=4\pi/3$.\cite{Huse,Kondev,Korshunov}
Assuming a small gap, $m_{\infty}=e^{-\frac{1}{2}\mathbf{Q}^2 \langle \mathbf{h}^2 \rangle} =e^{- \frac{8\pi c}{9 K} }$ and by using $m_{\infty} \sim 0.65$, we extract $K \sim 90t$.

\textit{Thermal effects}. We first discuss the thermal restoration of
the discrete symmetry of the effective model: the interface described
by (\ref{action}) undergoes a roughening transition in the
Kosterlitz-Thouless universality class (triggered by the condensation
of gluons) at a critical temperature usually obtained from the scaling
dimension of the locking potential, $2-|\mathbf{Q}_{\alpha}|^2 T/4 \pi K$,
leading to $T_{c}=3K/(2\pi)$.  For $T<T_{c}$ the system breaks the
discrete Z$_3$ symmetry with a $\sqrt{3} \times \sqrt{3}$ state, the
weathervane modes (gluons) are gapped and the defects (quarks) are
confined. For $T>T_{c}$ the Z$_3$ symmetry is restored, it is a
critical quark-gluon plasma with gapless gluons ($\eta=4/3$,\cite{Huse} hence a dynamical response $\chi''(\omega) \sim \omega^{\eta/2-1}$). For integer spins, we estimate
$T_{c}/J=0.16$ ($S=1$), $0.06$ ($S=2$), $0.02$ ($S=3$),
\textit{etc.} by using $K \sim 90 t$ and Eq.~\ref{t}. For half-integer
spins, since $t=0$ (at the one-loop order) and $t \ll t_6$ (with two-loop co-tunneling),\cite{CR} the system is immediately in its infinite temperature
phase and is critical. Could this deconfinement transition manifest
itself experimentally?  The transition should indeed occur in the corresponding spin-ice system with a discrete symmetry. In the model (\ref{H}) with SU(2)
symmetry, the Mermin-Wagner theorem ensures that no continuous
symmetry breaking will occur at finite temperature: the propagating
spin-waves induce a finite correlation length, $\xi_{2d} \sim
\exp(\rho/T)$,\cite{Chakra} which cuts off the infrared divergence of
the Kosterlitz-Thouless transition.\cite{u1} Strictly speaking
$T_{c}$ is a crossover temperature between two paramagnets. However,
the usual thermodynamic singularities will remain sharp because
$\xi_{2d} \sim \exp(JS^2/T_{c})$ is very large at $T=T_{c}$.

In conclusion, we argued that the Heisenberg kagome antiferromagnet
has long-range N\'eel order at $T=0$ for integer spins. Its magnetism
is unconventional due to an underlying emergent chromodynamics: there are color fluctuations and additional
low-energy excitations described as gapped ``gluons'' which may
condense at $T_c$ (``quark'' deconfinement).  For half-integer spins, $T_c=0$ at the lowest order. The difference between
half-integer and integer spins reflects the essential quantum nature
of the local modes.

We would like to thank P. Bruno, B. Canals, J. Jacobsen, K. Nguyen and T. Ziman for discussions.


\begin{thebibliography}{99}
\bibitem{ortho} The overlap of two states is at most $4^{-LS}$.
\bibitem{gauge} $[U_c,H_e]=0$ for all local operators, $U_c=\exp[i\alpha \sum_{\bigtriangleup} \lambda_{c,i}^z]$, where $c=3,8$ is the index of the Gell-Mann matrices.
\bibitem{u1}  If $H$ has a U(1) symmetry, below $T_c$, the state is critical, implying a second transition.\cite{Korshunov}
\end{thebibliography}
\end{document}